\documentclass{aastex}
\usepackage{emulateapj5}
\usepackage{apjfonts}

\newcommand{\swas}{{\it SWAS}}
\newcommand{\water}{H$_2$O}
\newcommand{\wateriso}{H$_2^{18}$O}
\newcommand{\ceio}{C$^{18}$O}
\newcommand{\thco}{$^{13}$CO}
\newcommand{\chhhcch}{CH$_3$CCH}
\newcommand{\hh}{H$_2$}

\journalinfo{{\rm Published in} The Astrophysical Journal, 539:L93--L96,
             {\rm 2000 August 20}}
\slugcomment{Received 1999 December 7; accepted 2000 June 20; published
             2000 August 16}

\shortauthors{EXTENDED WATER EMISSION IN ORION}
\shorttitle{SNELL ET AL.}

\begin{document}

\title{{\it Submillimeter Wave Astronomy Satellite} Observations of
       Extended Water Emission in Orion}

\author{R. L. Snell\altaffilmark{1},
J. E. Howe\altaffilmark{1},
M. L. N. Ashby\altaffilmark{2},
E. A. Bergin\altaffilmark{2},
G. Chin\altaffilmark{3},
N. R. Erickson\altaffilmark{1},
P. F. Goldsmith\altaffilmark{4},
M. Harwit\altaffilmark{5},
S. C. Kleiner\altaffilmark{2},
D. G. Koch\altaffilmark{6},
D. A. Neufeld\altaffilmark{7},
B. M. Patten\altaffilmark{2},
R. Plume\altaffilmark{2},
R. Schieder\altaffilmark{8},
J. R. Stauffer\altaffilmark{2},
V. Tolls\altaffilmark{2},
Z. Wang\altaffilmark{2},
G. Winnewisser\altaffilmark{8},
Y. F. Zhang\altaffilmark{2},\\
and G. J. Melnick\altaffilmark{2}}

\altaffiltext{1}{Department of Astronomy, University of Massachusetts,
                 Amherst, MA 01003}
\altaffiltext{2}{Harvard-Smithsonian Center for Astrophysics, 60 Garden Street,
                 Cambridge, MA 02138}
\altaffiltext{3}{NASA Goddard Spaceflight Center, Greenbelt, MD 20771}
\altaffiltext{4}{National Astronomy and Ionosphere Center, Department of
                 Astronomy, Cornell University, Space Sciences Building,
                 Ithaca, NY 14853-6801}
\altaffiltext{5}{511 H Street SW, Washington, DC 20024-2725; also Cornell
                 University}
\altaffiltext{6}{NASA Ames Research Center, Moffett Field, CA 94035}
\altaffiltext{7}{Department of Physics and Astronomy, Johns Hopkins University,
                 3400 North Charles Street, Baltimore, MD 21218}
\altaffiltext{8}{I. Physikalisches Institut, Universit\"{a}t zu K\"{o}ln,
                 Z\"{u}lpicher Strasse 77, D-50937 K\"{o}ln, Germany}

\begin{abstract}
We have used the {\it Submillimeter Wave Astronomy Satellite} to map
the ground-state $1_{10}\rightarrow1_{01}$ transition of ortho-\water\
at 557 GHz in the Orion molecular cloud.  \water\ emission was detected
in Orion over an angular extent of about 20\arcmin, or nearly 3 pc.
The water emission is relatively weak, with line widths (3--6 km s$^{-1}$)
and $V_{\rm LSR}$ velocities (9--11 km s$^{-1}$) consistent with an origin in
the cold gas of the molecular ridge.  We find that the ortho-\water\
abundance relative to H$_2$ in the extended gas in Orion varies
between 1 and 8$\times10^{-8}$, with an average of 3$\times10^{-8}$.
The absence of detectable narrow-line ortho-\wateriso\ emission is used
to set a $3\sigma$\ upper limit on the relative ortho-\water\ abundance
of 7$\times10^{-8}$.
\end{abstract}


\section{Introduction}

We present a map of the water emission from the central portion
of the Orion molecular cloud obtained with the {\it Submillimeter
Wave Astronomy Satellite} (\swas).  These results provide the first
definitive detection of water in the extended, cold molecular gas.
The Orion molecular cloud has been the focus of many studies of
water, and besides the strongly masing transition at 22 GHz, numerous
transitions of \water\ and \wateriso\ have been detected from the
ground \citep*{wat80,phi80,kui84, cer94,gen96,cer99}, from the {\it
Kuiper Airborne Observatory} \citep{zmu95,tim96}, and from the {\it
Infrared Space Observatory} \citep{har98, gon98, dis98}.  With the
exception of the broad \wateriso\ line emission observed toward BN/KL by
\citet{zmu95}, all of the transitions detected have upper states lying
from 200 to 700 K above the ground state and many are likely masing.
In most cases this emission arises from the hot gas associated with
BN/KL, however \cite{cer94} report spatially extended emission from the
$3_{13}\rightarrow 2_{20}$ transition of water at 183 GHz.

\swas\ observed the $1_{10} \rightarrow 1_{01}$ transitions of ortho-\water\
at 556.936 GHz and of ortho-\wateriso\ at 547.676 GHz.  These transitions
have upper state energies of only 27 K above the ortho-\water\ ground
state, and thus are unique probes of water in the cooler molecular gas.
The emission obtained with \swas\ toward BN/KL \citep{mel00a} is composed
of both a broad velocity feature associated with the shocked gas and a
narrow feature that likely arises from the quiescent gas in the Orion
molecular ridge.  The abundance of water relative to \hh\ in the shocked
gas is of order $4\times10^{-4}$, making water one of the most abundant
molecules in the hot gas. The focus of this Letter is the kinematically
narrow and spatially extended component of the water emission detected
in Orion.

\section{Observations and Results}

The observations reported here were acquired by \swas\ during the periods
from 1998 December to 1999 March and 1999 September to 1999 October.
The data were taken by alternately nodding the satellite from the source
to a reference position free of molecular emission.  Observations of the
$1_{10} \rightarrow 1_{ 01}$ transition of \water\ were obtained at 40
positions in a 5 by 8 grid spaced by 3\farcm2. The reference position
for this map was at $\alpha = 05^{\rm h} 35^{\rm m} 14\fs5$, $\delta$
= $-05\arcdeg  22\arcmin  37\arcsec$ (J2000).  Integration times were
typically 2 hr per position.  Besides the \wateriso\ spectrum reported
by \citet{mel00a} toward BN/KL, observations of \wateriso\ were also
obtained at a position 3\farcm2 south of the map reference position
where the \water\ spectrum showed a strong, narrow line.  The \swas\ beam
is elliptical, and at the frequency of the water transitions has angular
dimensions of $3\farcm3 \times 4\farcm5$.  The data were reduced using
the standard \swas\ pipeline described in \citet{mel00b}.  The data shown
in this paper are not corrected for the \swas\ main beam efficiency of 0.90.

We also used the Five College Radio Astronomy Observatory (FCRAO) 14~m
telescope to map an approximately $40\arcmin \times 130\arcmin$ region
in the $J=1 \rightarrow 0$ transition of \thco, a $18\arcmin \times
23\arcmin$ region in the $J=1 \rightarrow 0$ transition of \ceio, and
a $6\arcmin \times 6\arcmin$ region in the $J = 6 \rightarrow 5\ (K =
0,1,2,3,4)$ transitions of \chhhcch.  These data are used to provide an
estimate of the temperature and column density of the gas ridge for our
analysis of the water emission.

\clearpage

\centerline{\includegraphics[width=0.92\hsize,clip]{orion_h2o_fig1.ps}}
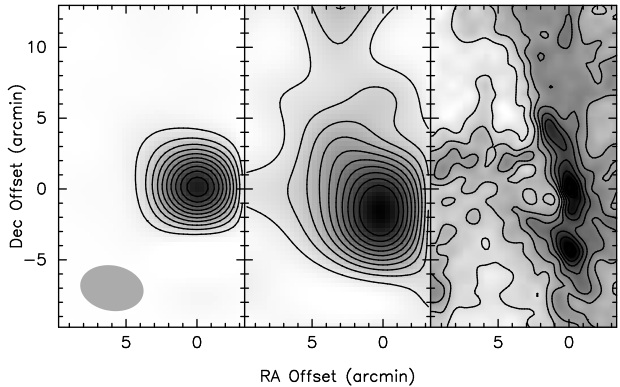
\figcaption{The left-hand panel shows a map of the integrated intensity
of the 557 GHz \water\ line in Orion.  The map contours ($\int T_A^*dv$)
are between 10 and 100 K km s$^{-1}$ in steps of 10 K km s$^{-1}$.
The integrated intensity is dominated by the broad emission associated
with BN/KL.  The middle panel shows a map of the integrated intensity
of just the narrow water line component in Orion. Contours ($\int
T_A^*dv$) in this map are between 1 and 10 K km s$^{-1}$ in steps
of 1 K km s$^{-1}$. The right hand panel is a map of the integrated
intensity of the \thco\ $J=1\rightarrow0$ emission obtained at FCRAO.
Contours are between 5 and 45 K km s$^{-1}$ in steps of 5 K km s$^{-1}$.
The offsets are relative to $\alpha = 05^{\rm h}35^{\rm m}14\farcs5$,
$\delta = -05\arcdeg22\arcmin37\arcsec$ (J2000).  The half-power size and
average position angle of the \swas\ beam is also shown in the first panel.
\label{fig1}}
\vspace{0.9\baselineskip}

\water\ emission was detected in 23 of the 40 positions observed.
The left-hand panel of Figure 1 shows the integrated intensity map of
the \water\ emission, which is dominated by the broad line width emission
associated with BN/KL.  Because of the large and slightly elliptical main
beam of \swas, this broad emission is seen in five of the map positions.
Outside of these five positions the emission is narrow with line widths
between 3 and 6 km s$^{-1}$.  Figure 2 provides examples of \water\
spectra obtained at three positions in the cloud.  We have fitted the
positions showing broad emission with two Gaussian components and include
only the integrated intensity of the narrow component in the map shown
in the middle panel of Figure 1.  The narrow line width water emission
is elongated north-south, and extends 12\farcm8 north of BN/KL to OMC-2
and 6\farcm4 south of BN/KL.  The velocity of the narrow line width
emission varies from $V_{\rm LSR} = 8.7$ to 11.6 km s$^{-1}$, with a trend
of increasing velocity from south to north.  The line widths and velocity
of the \water\ emission agree well with the emission associated with the
extended molecular ridge and molecular bar.  Weak \water\ emission also
extends 10\arcmin\ east of the molecular ridge where weaker emission is
detected in both \thco\ and CN \citep*{rod98}.

\section{Water Abundance}

The $1_{10} \rightarrow 1_{01}$ transition of \water\ has a high critical
density ($\thicksim 4\times10^{8}$ cm$^{-3}$) and is expected to have
a high optical depth even for a relatively small water abundance.
Thus, trapping plays an important role in the excitation of this
transition. For large optical depths the ``effective critical density''
is approximately $A/(C\tau_o)$, where $C$ is the collisional de-excitation
rate coefficient, $\tau_o$ is the line center optical depth, and $A$
is the spontaneous emission rate.  In the low-collision rate limit,
where the density is much smaller than the effective critical density,
\citet{lin77} and \citet{wan91} derived a simple analytical expression
for the antenna temperature in a two-level system.  In this limit,
collisional excitation always results in a photon that escapes the cloud,
even though it may be repeatedly absorbed and reemitted.  Thus the gas
can be optically thick but effectively thin.

\centerline{\includegraphics[width=0.92\hsize,clip]{orion_h2o_fig2.ps}}
\figcaption{Spectra of \water\ and \wateriso\ obtained with \swas.
In order from the top are spectra of \water\ 3\farcm2 south of BN/KL,
\wateriso\ 3\farcm2 south of BN/KL (note that the spectrum has been
scaled by a factor of 10), \water\ 6\farcm4 east of BN/KL, and \water\
3\farcm2 east and 12\farcm8 north of BN/KL near the position of OMC-2.
\label{fig2}}
\vspace{0.9\baselineskip}

Applying this approximation to water, gives the following expression
for the integrated antenna temperature for the 557 GHz water line:

\begin{equation}
\int T_R dv = Cn_{\rm H_2} {c^3\over2\nu^3k} N(\hbox{o-H$_2$O}) {h\nu\over4\pi}
\exp(-h\nu/kT_K).
\end{equation}

\noindent
Note that in this limit the integrated intensity increases linearly with
increasing H$_2$ density, $n_{\rm H_2}$, and ortho-water column density,
$N$(o-\water), even if the line center optical depth is large.  We have
evaluated equation (1) using the collision rate coefficients for para-
and ortho-\hh\ from \citet*{phi96} for a temperature of 40 K, assuming the
ratio of ortho- to para-\hh\ is in LTE (implying an ortho-to-para ratio
of 0.13).  The ortho-water fractional abundance, $x$(o-H$_2$O), is thus

\begin{equation}
x(\hbox{o-H$_2$O}) = 2.5\times10^{19} {\int T_R dv\over N({\rm H_2})n_{\rm H_2}}, 
\end{equation}

\noindent
where the integrated intensity of the water line has units of
Kelvins kilometers per second, density has units per cubic centimeter,
and column density has units per square centimeter.  If we assume that
the molecular ridge has $T =$ 40 K, $n_{\rm H_2} = 1\times10^6$ cm$^{-3}$,
and $N$(\hh) = $5\times10^{22}$ cm$^{-2}$, then for a typical main beam
corrected water integrated intensity in the ridge of 10 K km s$^{-1}$,
the relative abundance of ortho-\water\ is $5\times10^{-9}$.  Although the
line center optical depth is over 10, the emission is still well within
the low-collision rate limit, so the emission is effectively thin.
This estimate ignores beam dilution which could be significant in the
3\farcm3 $\times$ 4\farcm5 main beam of \swas.

A detailed model of the temperature, density, column density, and
velocity dispersion for the Orion molecular gas is needed to model the
water excitation and beam filling factor to derive accurate estimates
of the water abundance.  We have used our observations of \chhhcch\
to derive the temperature along the molecular ridge in Orion using
the method described by \citet{ber94}.  We find temperatures varying
from 50 K toward the center of the ridge dropping to about 25 K at the
boundary of the \chhhcch\ emission, consistent with temperatures found
by \citet{ber94}.  We have assumed a temperature of 25 K outside of the
area of detectable \chhhcch\ emission.  The H$_2$ gas column density is
derived from our observations of \thco\ assuming LTE, the temperatures
described above, and a \thco/\hh\ abundance ratio of $1.5\times10^{-6}$.
The column density is derived on a 44\arcsec\ grid defined by the
\thco\ sampling.  The velocity dispersion of the gas along each line of
sight is determined from the \thco\ line width.  We have assumed that the
density of the gas along the molecular ridge near BN/KL is $1\times10^6$
cm$^{-3}$ \citep*{bat83,ber96,rod98}.  The extension of the ridge north
to OMC-2 is assumed to have a density of $3\times10^{5}$ cm$^{-3}$
based on the analysis of \citet{bat83}, and the density of the diffuse
emission to the east of BN/KL is assumed to be $3\times10^5$ cm$^{-3}$
based on the CN analysis of \citet{rod98}.  We have assumed that all of
the gas column density toward the Orion core arises in the dense gas,
consistent with the results found by \citep{ber96}.

Once the temperature, density, velocity dispersion, and \hh\ column
density are fixed, the strength of the water emission is determined only
by its abundance relative to \hh.  We model the water emission using
a statistical equilibrium code that uses the large velocity gradient
approximation to account for radiation trapping.  Collisional rate
coefficients are taken from \citet{phi96} using both the para- and
ortho-\hh\ rates and assuming that the ratio of ortho- to para-\hh\ is
in LTE\footnote{ We note that assuming an ortho- to para-\hh\ ratio of 3
(the high temperature limit), the collision rates would be approximately 5
times larger and the derived water abundance correspondingly smaller.}.
We include the 5 lowest levels of ortho-water in our calculations.
We proceed by assuming a water abundance, computing the water emission in
each of the 44\arcsec\ regions, convolving the predicted emission with the
\swas\ beam, and then varying the abundance until there is agreement between
the integrated intensity predicted by the model and the observations.
We assume the \water\ abundance is constant across the \swas\ beam.  Using
this technique, we find that for the 23 positions where \water\ emission
was detected that the average abundance of ortho-\water\ relative to \hh\
is $3\times10^{-8}$, with variations between 1 and 8$\times10^{-8}$.
For the narrow velocity component detected toward BN/KL, our results
agree with \citet{mel00a}.  The largest \water\ abundances are found
east of the molecular ridge near the molecular bar and H~{\sc II} region.

We can also use the \wateriso\ spectrum obtained 3\farcm2 south of BN/KL
(see Fig.\ 2) to estimate the water abundance.  The \wateriso\ spectrum
may show evidence for a weak broad line width component similar to that
detected toward BN/KL \citep{mel00a}.  For the \wateriso\ observation the
long axis of the \swas\ beam was oriented nearly north/south and explains
why the broad \wateriso\ emission toward BN/KL may be weakly detected at
this position.  We have fitted the \wateriso\ spectrum with two components,
one matching the line width and velocity of the broad velocity component
seen toward BN/KL and a second narrow-line component with a width (5.65
km s$^{-1}$) and velocity (9.0 km s$^{-1}$) derived from the \water\
line detected in this direction.  No evidence for a narrow-line velocity
component was found in the \wateriso\ spectrum, similar to the results
found by \citet{mel00a} toward BN/KL.  Using the same model as before,
the \wateriso\ data sets a $3\sigma$\ upper limit of $7\times10^{-8}$
for the relative abundance of ortho-\water\ 3\farcm2 south of BN/KL,
assuming a ratio of H$_2$O/H$_2^{18}$O of 500.  This result is consistent
with the abundance of $4\times10^{-8}$ derived from the \water\ analysis.

Our modeling has ignored several potentially important effects: (1)
pumping by the far-infrared radiation field of the cloud, (2) line
scattering in a thin layer of gas surrounding the cloud core, and (3)
clumpy cloud structure.  The importance of radiation pumping can be
easily estimated by computing the excitation of a molecule immersed
in a dilute blackbody radiation field derived from the measurements
toward BN/KL \citep*{wer76,kee82}.  Even using the strong radiation
field present within 1\arcmin\ of BN/KL, the fractional population in
the upper rotational states (2$_{12}$ and 2$_{21}$) of water will be
small, and thus ignoring this radiation will have a negligible effect
on our estimates of the water abundance.  Toward BN/KL, \citet{mel00a}
have more carefully modeled the effect of dust on \water\ excitation and
confirm that for the extended emission ignoring dust has only a small
impact on the derived \water\ abundance.

Line scattering is of even greater concern.  Because of the low
excitation, but high optical depths predicted for this water transition,
photons collisionally produced in the ridge may be absorbed and reemitted
by an extended halo of low column density gas.  In this case the abundance
of water derived for the molecular ridge may be underestimated and the
abundance of water in the extended structure of the Orion cloud may be
overestimated.  The importance of this effect for HCO$^+$ was investigated
by \citet{cer87}.  The water line shape provides some constraints on
this process.  The water lines show no evidence for self-absorption as
would be expected if absorption in a surrounding envelope is important.
While absorption followed by emission is plausibly occurring, the photons
would emerge in the line wings after a number of scatterings and most
of the scattered photons would probably still be within the \swas\ beam.
The only exception would be if the absorbing envelope were spatially
well separated from the emitting gas.  Nevertheless, even if scattering
were having a large impact on the \water\ emission, it is not important
for \wateriso\, and therefore the abundance limit established from the
\wateriso\ observation is valid.

The Orion cloud core may also have structure on scales smaller than
modeled.  If there is significant structure on scales smaller than
44\arcsec\, we may underestimate the optical depth of water in the high
column density regions.  This in turn, could cause an underestimate of
the collisional de-excitation rate and consequently an underestimate of
the water abundance.  This might be important near BN/KL; however, it is
unlikely to be important in the more extended water emission regions
where the emission is considerably weaker and, even with significant
unresolved structure, it is unlikely the column density is sufficiently
large for collisional de-excitation to be important.  Regardless of its
impact on \water\, this effect is unimportant for \wateriso\, and our
upper limit on the water abundance based on \wateriso\ is unaffected.
However it is important to stress that the \water\ abundances that we
quote are averages over the \swas\ beam, it is possible that there are
significant abundance variations within the beam.

\section{Discussion and Summary}

\swas\ has made the first detection of water emission from the cold,
quiescent gas in the Orion molecular cloud.  The water emission seen by
\swas\ has line widths and velocities characteristic of the emission from
the molecular ridge. The average abundance of ortho-\water\ in the cold
gas is $3\times10^{-8}$, with variations between 1 and 8$\times10^{-8}$.
As discussed earlier, the abundance derived from the \water\ emission can
be compromised by several physical processes, however a firm  $3\sigma$\
upper limit on the relative water abundance 3\farcm2 south of BN/KL of
$7\times10^{-8}$ is established from the \wateriso\ observation.

Uncertainties in the water abundance arise entirely from systematic
errors associated with uncertainties in our estimates of the density and
gas column density.  For instance, if a significant fraction of the
gas column density is associated with much lower density gas that we
assumed in our model, then our estimate of the ortho-water abundance will
be underestimated.  Specifically, if one-half of the gas column density
originated in low-density gas, our estimate of the water abundance will be
too low by approximately a factor of two.  Since \citet{ber96} concluded
that it is unlikely that the low density gas contributes significantly
to the total gas column density toward the Orion core, we believe that
our estimates of the water abundance cannot be greatly in error.

\citet{cer94} estimated the water abundance in Orion based on
their observations of the  $3_{13} \rightarrow 2_{20}$ transition
of para-\water\ at 183 GHz.  Although this masing transition arises
from an upper state lying 200 K above the ground state, the emission is
spatially extended and, away from BN/KL, has relatively narrow line widths.
\citet{cer94}  estimate a relative water abundance of at least 10$^{-5}$.
However, if the abundance of water in the cold molecular gas were
as large as they estimate, then \wateriso\ emission should have been
readily detected by \swas\ and be of comparable strength to the detected
\water\ emission.  In addition, the \water\ line observed by \swas\
should be between 5 and 100 times stronger than observed, depending on
position. The 183 GHz line must trace a warmer, denser component of the
gas than the 557 GHz line - as suggested by the anomalous line velocity
reported by \citet{cer94}  - so that the high abundance estimate does
not apply to the bulk of the cloud material, which is at low temperature.

The water abundance we derive for the quiescent gas in Orion is many
orders of magnitude smaller than that derived for the hot gas in Orion and
other massive star forming regions \citep{mel00a,cer99,dis98b}.  Chemical
models \citep*{ber98} predict equilibrium gas-phase water abundances of
order $5\times10^{-7}$ in the quiescent gas, similar to the results found
for the extended envelope surrounding SgrB2 \citep{zmu95}, but about
10 times larger than what we estimate for Orion.  Further discussion of
the chemical implications of these results is presented in \citet{ber00}.

\acknowledgements

This work was supported by NASA's \swas\ contract NAS5-30702 and NSF
grant AST 97-25951 to the Five College Radio Astronomy Observatory.
R. Schieder \& G. Winnewisser would like to acknowledge the generous
support provided by the DLR through grants 50 0090 090 and 50 0099 011.

\end{document}